\documentclass[12pt]{article}
\usepackage{epsfig}
\newcommand{\nc}{\newcommand}
\newcommand{\as}{\alpha_s}
\def\eqn#1{eq.~(\ref{#1})}

\def\sec#1{section~{\ref{#1}}}

\nc{\fig}[1]{Fig.~\ref{fig:#1}}
\nc{\figs}[2]{Figs.~\ref{fig:#1} and \ref{fig:#2}}
\renewcommand{\slash}[1]{/\kern-7pt #1}
\nc{\beq}{\begin{equation}}
\nc{\eeq}{\end{equation}}
\nc{\bea}{\begin{eqnarray}}
\nc{\eea}{\end{eqnarray}}
\nc{\beas}{\begin{eqnarray*}}
\nc{\eeas}{\end{eqnarray*}}
\nc{\nn}{\nonumber}
\nc{\gsim}{\raisebox{.2em}{$\rlap{\raisebox{-.5em}{$\;\sim$}}>\,$}}
\nc{\lsim}{\raisebox{.2em}{$\rlap{\raisebox{-.5em}{$\;\sim$}}<\,$}}
\nc{\eps}{\epsilon}
\def\Ord{{\cal O}}
\def\e{\epsilon}
\def\eps{\epsilon}
\def\qb{{\bar q}}
\def\lr{\leftrightarrow}
\def\tree{{\rm tree}}
\def\spa#1.#2{\left\langle#1 \hskip .15 mm #2\right\rangle}
\def\spb#1.#2{\left[#1 \hskip .15 mm #2\right]}
\def\spaa#1.#2.#3{\langle\mskip-1mu{#1} 
                  | #2 | {#3}\mskip-1mu\rangle}
\def\spab#1.#2.#3{\langle\mskip-1mu{#1} 
                  | #2 | {#3}\mskip-1mu\rangle}
\def\cg{c_\Gamma}
\def\sstw{\sin^2\theta_W}
\def\Br{{\rm Br}}
  \newcommand{\ccaption}[2]{
    \begin{center}
    \parbox{0.85\textwidth}{
      \bigskip
      \caption[#1]{\small{#2}}
      }
    \end{center}
    }
\newbox\charbox
\newbox\slabox
\def\s#1{{      
        \setbox\charbox=\hbox{$#1$}
        \setbox\slabox=\hbox{$/$}
        \dimen\charbox=\ht\slabox
        \advance\dimen\charbox by -\dp\slabox
        \advance\dimen\charbox by -\ht\charbox
        \advance\dimen\charbox by \dp\charbox
        \divide\dimen\charbox by 2
        \raise-\dimen\charbox\hbox to \wd\charbox{\hss/\hss}
        \llap{$#1$}
}}

\begin{document}

\begin{titlepage}
\begin{flushright}
SLAC--PUB--8188 \\
ETH-TH/99-10  \\
DTP/99/72 \\
hep-ph/9907305 \\ 
July 1999 
\end{flushright}
\vskip 1.2cm
\begin{center}
\boldmath
{\Large\bf Vector Boson Pair Production} 
\vskip .2cm
{\Large\bf in Hadronic Collisions at  ${\cal O}(\alpha_s)$: } 
\vskip .2cm
{\Large\bf Lepton Correlations and Anomalous Couplings}

\unboldmath 
\vskip 1.2cm
{\sc L. Dixon}
\vskip .3cm
{\it Stanford Linear Accelerator Center, Stanford University,
Stanford, CA 94309, USA }
\vskip .7cm
{\sc Z. Kunszt}
\vskip .3cm
{\it Theoretical Physics, ETH Z\"urich, Switzerland }
\vskip 0.7cm
{\sc A. Signer}
\vskip .3cm
{\it Department of Physics, University of Durham,
Durham DH1 3LE, England}
\vskip 2.cm

\end{center}

\begin{abstract}
\noindent 
We present cross sections for production of electroweak vector boson
pairs, $WW$, $WZ$ and $ZZ$, in $p\bar{p}$ and $pp$ collisions, at
next-to-leading order in $\alpha_s$.  We treat the leptonic decays
of the bosons in the narrow-width approximation, but retain all spin
information via decay angle correlations.  We also include the effects
of $WWZ$ and $WW\gamma$ anomalous couplings.
\end{abstract}

\begin{center}
\vskip 1.5cm
{\it Submitted to Physical Review D}
\end{center}

\end{titlepage}

\section{Introduction}

At the core of the electroweak Standard Model is its invariance under
the nonabelian gauge group $SU(2) \times U(1)$.  Many aspects of this
gauge structure, such as vector boson masses and couplings to
fermions, have already been tested with high precision in a variety of
experiments.  However, the nonabelian self-interactions of vector
bosons --- in particular, triple gauge-boson couplings --- are just
beginning to be studied directly, via vector-boson pair production in
$e^+e^-$ annihilation at LEP2 at CERN, and in $p\bar{p}$ collisions at
Run I of the Fermilab Tevatron.  Although thousands of $W^+W^-$ pairs
have been collected at LEP2, they have all been produced at relatively
modest values of the pair invariant mass, $M_{WW} \lsim 200$ GeV.  On
the other hand, if there are anomalous (non-Standard Model)
vector-boson self-couplings, their effects are expected to grow with
invariant mass, so it is useful to study vector-boson pair production
at the highest possible energies.  Vector boson pairs also provide a
background for other types of physics.  If the Higgs boson is heavy
enough it will decay primarily into $W^+ W^-$ and $ZZ$
pairs~\cite{EHLQ}.  Exotic Higgs sectors can have substantial
branching ratios for charged Higgs bosons to decay to $WZ$
pairs~\cite{IUK}.  Leptonically decaying $WZ$ pairs, $W^+Z \to
\ell^+\nu \ell^{\prime +}\ell^{\prime -}$, in which the
negatively-charged lepton is lost, form a background to a signal for
strong $WW$ scattering associated with the mode $W^+W^+ \to \ell^+ \nu
\ell^{\prime +} \nu^\prime$~\cite{CK}.  Finally, a prime signal for
supersymmetry at hadron colliders is the production of three charged
leptons and missing transverse momentum~\cite{TriLepton}; a background
for this process is the production of a $W$ plus a (virtual) $Z$ or
$\gamma$.

In the near future, hadron colliders will be the primary source of
vector boson pairs with large invariant mass.  Run II of the upgraded
Tevatron should yield a data set roughly 20 times larger than that
from Run I, including 100--200 leptonically decaying $W^+W^-$ pairs.
The Large Hadron Collider (LHC) at CERN promises to increase the
sample by another factor of 50 beyond Run II.  With this increase in
statistics, refined Standard Model predictions are essential.  The
leading QCD corrections ($\Ord(\alpha_s)$) are significant (generally
of order tens of percent), and hence are required to get a precise
estimate of the overall production cross section.  Also, experiments
do not detect vector bosons, but only those leptonic decay products
that fall within the experimental acceptance.\footnote{Modes in which
one of the vector bosons decays hadronically have been studied at the
Tevatron, but at Standard Model levels these events are hard to
separate from the QCD production of a vector boson plus
jets~\cite{LeptonPlusJets}.}  Spin correlations between vector bosons
are reflected in kinematic distributions of leptonic momenta, which in
turn influence the number of events surviving experimental cuts.  In
order to properly take into account the effects of cuts on the cross
section, as well as to study the more detailed (lepton) kinematic
distributions permitted by higher statistics, it is important to treat
the vector boson decays properly, including all spin correlations.

Hadronic production of vector boson pairs in the Standard Model has
already been studied extensively.  The Born-level, or leading-order (LO)
cross sections for $W^+ W^-$, $W^\pm Z$ and $ZZ$ pair production via quark
annihilation were computed twenty years ago~\cite{TreePair}.  These
cross sections were evaluated by treating the $W$ and $Z$ as stable
particles and summing over their polarization states, using completeness
relations to simplify the sum; thus spin and decay correlations were
neglected.  For the spin-summed production cross section, the
next-to-leading order (NLO, or $\Ord(\alpha_s)$) QCD corrections were
obtained for $W^+W^-$ final states in refs.~\cite{WWOhn,WWit}, for 
$W^\pm Z$ in refs.~\cite{WZOhn,WZit}, and for $ZZ$ in 
refs.~\cite{ZZOhn,ZZit}.

The simplest way to include the effects of vector-boson spin and decay
cor\-re\-la\-tions is to compute directly the matrix elements for the
production of the four final-state fer\-mions.  In the narrow-width
approximation, only `doubly-resonant' Feynman diagrams have to be
considered --- the same class of diagrams that gives rise to the
on-shell spin-summed cross section.  Because both the outgoing
fermions and the initial-state partons are essentially massless, and
because their couplings to vector bosons are chiral, it is very
convenient to use a helicity basis for the fermions.  The tree-level
helicity amplitudes for massive vector-boson pair production and decay
into leptons were first computed in ref.~\cite{GK}, which also
demonstrated the significance of decay-angle correlations.  This same
approach was carried out at order $\alpha_s$ in ref.~\cite{OhnSpin}.
At $\Ord(\alpha_s)$, there are real corrections, consisting of tree
graphs with an additional gluon in either the initial or final state,
and virtual corrections, consisting of one-loop amplitudes that
interfere with the Born amplitude.  However, the full one-loop
amplitudes including leptonic decays were unavailable until recently,
so ref.~\cite{OhnSpin} included decay correlations everywhere except
for the finite part of the virtual contribution, for which spin-summed
formulae were used.\footnote{The virtual corrections can be divided
into terms with poles in $\e$, the parameter of dimensional
regularization, plus residual finite terms.  The pole terms have a
universal form and cancel against infrared divergences in the real
corrections, so if the real corrections include decay correlations,
then the virtual pole terms must also, in order to get a finite
answer.}

In this paper, we present $\Ord(\alpha_s)$ results for the hadronic
production of $W^+W^-$, $WZ$ and $ZZ$ pairs, including the full lepton
decay correlations in the narrow-width approximation.  We rely on
ref.~\cite{DKS} for all the required matrix elements, in particular the
virtual one-loop amplitudes for $q\bar{q}' \to V_1V_2 \to 4$ leptons.
In order to cancel the infrared divergences in the phase-space
integrations for the real corrections, we implement the general 
subtraction method discussed in ref.~\cite{FKS}.  This method allows
the computation of distributions of arbitrary (infrared-safe) observables.

Recently, an update to vector boson pair production has been presented
in ref.~\cite{CE}.  The corresponding Monte Carlo program, {\tt MCFM},
relies on the same amplitudes~\cite{DKS} and, therefore, also includes
all spin correlations exactly to next-to-leading order in $\as$.  {\tt
MCFM} is more complete than the program described here, in the sense
that the narrow-width approximation is not assumed, and
singly-resonant diagrams are also included.  These additions are
expected to shift the resonance-dominated di-vector boson cross
sections by the order of several percent. Their effects are obviously
much bigger in the off-resonant regions important for studies of
Standard Model backgrounds to new physics.

In the present paper, we first compute the di-vector boson cross sections
in the Standard Model, both without and with a realistic set of
experimental cuts.  For $W^+W^-$ production, a jet veto is used by 
experimentalists to suppress backgrounds; we study the effect of this veto
on the size of the cross section and its renormalization/factorization 
scale dependence.

Di-boson amplitudes in the Standard Model have interesting angular
dependences.  For example, at Born level, there is an exact radiation zero
in the partonic process $q_1\bar{q}_2 \to W^\pm\gamma$ at $\cos\theta =
(Q_1+Q_2)/(Q_1-Q_2)$, where $\theta$ is the scattering angle of the $W$
with respect to the direction of the quark $q_1$, and $Q_{1,2}$ are the
quark electric charges~\cite{TreePair}.  Similarly, it has been shown 
that there is a an approximate zero for $q_1\bar{q}_2 \to W^\pm Z$ at 
$\cos\theta = (g^-_1 + g^-_2)/(g^-_1 - g^-_2)$, where $g^-_{1,2}$ are 
the left-handed couplings of the $Z$ boson to the quarks~\cite{BHOWZZero}.
The exact tree-level zero for $W\gamma$ is filled in somewhat by 
QCD radiative corrections, and also by the kinematic ambiguity
associated with the undetected neutrino in 
$W^+\gamma \to \ell^+ \nu \gamma$.   Still it produces a dip in the 
distribution of a related variable, the rapidity difference 
$y_\gamma-y_{\ell^+}$, which should be visible at Run II~\cite{BEL}.  

Here we study the QCD corrections to the approximate $WZ$
radiation zero.  Using on-shell, spin-summed cross sections,
ref.~\cite{WZit} computed the distribution in the rapidity difference
between the $W$ and $Z$ bosons, $\Delta y_{WZ} = |y_W - y_Z|$, which is a
boost-invariant surrogate for the center-of-mass scattering angle
$\theta$.  It was found that a dip in the $\Delta y_{WZ}$ distribution
persists at $\Ord(\alpha_s)$, although the dip is less pronounced than at
Born level.  Since the rapidity of the $W$ cannot be determined on an
event-by-event basis, we study a quantity related to $\Delta y_{WZ}$, but
constructed purely out of charged lepton variables, and find that a dip
(or at least a shoulder) is still present.  However, because of the much
lower $WZ$ cross section, this measurement is considerably more 
challenging than the $W\gamma$ case, and will probably have to wait for 
the LHC.

Various types of TeV-scale physics may modify vector-boson
self-interactions.  Without a precise knowledge of the new physics,
one often parameterizes the modifications using anomalous coupling
coefficients.  We shall consider anomalous contributions to the $W^+
W^- Z$ and $W^+ W^- \gamma$ triple gauge vertices, and their effect on
various distributions in $WW$ and $WZ$ production.  Similar studies
have already been carried out at order $\alpha_s$ and including spin
correlations everywhere except in the finite virtual contributions,
for both $WW$ production~\cite{BHOWW} and $WZ$
production~\cite{BHOWZ}.  Here we include as well the spin correlation
effects from the finite virtual contributions.  This requires matrix
elements beyond those in ref.~\cite{DKS}; however, the new matrix
elements are trivial by comparison.

The remainder of the paper is organized as follows.  After outlining
the computational techniques in \sec{ComputationSection}, we present
results for the Standard Model production of $WW$, $WZ$ and $ZZ$ pairs
in \sec{SMSection}.  We first present total cross sections for all
three channels, without and then with a set of realistic kinematic
cuts on the leptons.  We consider both $p\bar{p}$ collisions at
$\sqrt{s} = 2$ TeV (corresponding to Run II of the Tevatron), and $pp$
collisions at $\sqrt{s} = 14$ TeV (the LHC).  We discuss the
dependence of the $WW$ cross section on a common renormalization and
factorization scale, with and without a jet veto.  For the $WZ$
channel, we study the approximate radiation zero before and after QCD
corrections.

In \sec{AnomalousSection} we introduce anomalous $W^+W^-Z$ and 
$W^+W^-\gamma$ couplings, and compute their effect on the matrix elements
in the narrow-width approximation through $\Ord(\alpha_s)$.  We then study
the effects of these couplings on a double-binned transverse energy 
distribution for the pair of charged leptons in $W^+W^-$ production
followed by leptonic decays.  Finally, in \sec{ConclusionsSection} we 
present our conclusions.

\section{Computation}
\label{ComputationSection}

It is straightforward to implement the helicity amplitudes presented
in ref.~\cite{DKS}, and those including anomalous couplings (see
section~\ref{AnomalousSection}), in a Monte Carlo program.  The
tree-level and one-loop amplitudes are computed as complex numbers and
the squaring, as well as the sum over helicity configurations, is done
numerically. In order to cancel singularities between the real and
virtual parts analytically, we use the general version of the
subtraction method~\cite{ERT} as presented in ref.~\cite{FKS}. Our
code is flexible enough to compute arbitrary infrared-safe quantities,
apply arbitrary cuts, and add any parton distribution
easily.\footnote{We wrote two independent programs, one in Fortran~77
and one in Fortran~90, both of which are available upon request.}

In this paper we will present results for the Tevatron Run II and the LHC.
The former term refers to $p\bar{p}$ scattering at $\sqrt{s}=2$~TeV,
whereas the latter stands for $p p$ scattering at $\sqrt{s}=14$~TeV.  Most
of our results will be presented with some `standard cuts' which are
defined as follows:  We make a transverse momentum cut of $p_T>20$~GeV 
for all charged leptons.  The event is required to have a minimum
missing transverse momentum $p_T^{\rm miss}$, which is carried off by the
neutrino(s).  We require $p_T^{\rm miss}>25$~GeV in the
case of $W$-pair production and $p_T^{\rm miss}>20$~GeV in the case of
$W^\pm Z$ production.  No $p_T^{\rm miss}$ cut is applied for $Z$-pair
production.   Finally, we apply some collider-dependent rapidity cuts for
the charged leptons.  For the Tevatron we require $|\eta| < 1.5$, whereas
for the LHC $|\eta| < 2.5$.

In all results presented in this paper, we assume that
the vector bosons always decay leptonically, i.e. the proper branching 
ratios of the vector boson decays into leptons $\Br(V\to f \bar{f}')$ 
are {\it not} included.  Obviously, these branching ratios depend on 
which final-state charged leptons are included in the analysis 
(electrons, muons, or both).  They can easily be added at any stage, 
using
\bea \Br(Z\to e^+ e^-) =  \Br(Z\to \mu^+ \mu^-) &=& 3.37\%, \nn 
\\ \sum_{i=e,\mu,\tau} \Br(Z\to\nu_i\bar{\nu_i}) &=& 20.1\%, \label{branching}
\\ \Br(W^+\to e^+\nu_e) =\ \Br(W^+\to \mu^+\nu_\mu) 
   &=& 10.8\%. \nn 
\eea
These ratios implicitly incorporate QCD corrections to 
the hadronic decay widths of the $W$ and $Z$.

We use two different parton distributions, MRST(ft08a)
\cite{mrst} and CTEQ(4M) \cite{cteq}, which we refer to simply as
MRST and CTEQ. For both the leading and next-to-leading order results
we shall use the same parton distributions (which have been obtained by 
a fit at next-to-leading order in $\as$). The strong coupling constant is
evaluated using 
\bea 
\as(\mu) &=& { \as(M_Z) \over w} \left( 1 -
\frac{\as(M_Z)}{\pi} {\beta_1\over\beta_0} {\ln(w)\over w} \right) \,
, \nn \\ 
\label{eq:twoloopalpha}
w &=& 1 - \beta_0 \frac{\as(M_Z)}{\pi} \ln\biggl({M_Z\over\mu}\biggr) \, , 
\eea
with $\beta_0 = \frac{1}{2} (\frac{11}{3}C_A - \frac{2}{3} N_f)$,
$\beta_1 = \frac{1}{4}
(\frac{17}{3}C_A^2-(\frac{5}{3}C_A+C_F)N_f)$, $C_A = N_c$, $C_F =
(N_c^2-1)/(2 N_c)$.  The value of $\as(M_Z)$ is set equal to the 
value given in the respective parton distribution fit. Thus, we
take $\as(M_Z) = 0.1175$ for MRST and $\as(M_Z) = 0.116$ for CTEQ. In
all computations we have set the renormalization and factorization
scales equal: $\mu_R = \mu_F \equiv \mu$.

The masses of the vector bosons have been set to $M_Z=91.187$~GeV and
$M_W=80.33$~GeV.  As for the coupling constants $\alpha$ and
$\sin^2\theta_W$, we choose them in the spirit of the 
``improved Born approximation''~\cite{IBA1,IBA2} for $W$ pair production at 
LEP2.  We do not explicitly include any QED or electroweak 
radiative corrections.  However, we take into account the top-quark-enhanced
corrections to the relation between $M_Z$, $M_W$ and $\sin^2\theta_W$,
where the latter is defined as an effective coupling in a high-energy
process, by using the definition~\cite{IBA2}
\beq
   \sin^2\theta_W \equiv { \pi \alpha(M_Z) \over \sqrt{2} G_F M_W^2 } \,,
\eeq
where $G_F = 1.16639 \times 10^{-5}$~GeV$^{-2}$ is the Fermi constant
and $\alpha(\mu)$ the running QED coupling.  For our numerical results 
we use $\alpha=\alpha(M_Z)=1/128$ and $\sin^2\theta_W=0.230$.  

The programs have been set up to allow for arbitrary values in the
entries of the Cabibbo-Kobayashi-Maskawa (CKM) mixing matrix that do
not depend on the top quark.  For the numerical results we take
$|V_{ud}|=|V_{cs}|=0.975$ and $|V_{us}|=|V_{cd}|=0.222$. 

Finally we mention that we do not properly include processes where top
quarks are involved. The amplitudes presented in ref.~\cite{DKS}
assume massless quarks, an approximation that is certainly not
justified in the case of the top quark. The fact that we nevertheless
include the $t$-channel exchange of the top quark (with
$|V_{td}|=|V_{ts}|=0$ and $|V_{tb}|=1$) for $W$-pair production,
therefore, results in an error. Fortunately, these processes are
suppressed for energy scales that are not too large, either by small
CKM matrix elements or by the small $b$ quark distribution
function. Indeed, we checked that the contribution of the subprocess
$b \bar{b} \to W^+ W^-$ (treating the top as massless) is completely
negligible for Run~II while it is of the order of 2\% for the LHC.
Furthermore, in the case of $W^\pm Z$ production we did not include
the process $ b g \to W^- Z t $. This process is present at
next-to-leading order but is strongly suppressed by the large top
quark mass, as well as the small $b$ quark distribution function.


\section{Standard Model Results}
\label{SMSection}

\subsection{Total Cross Sections}

The total cross sections for NLO vector-boson pair production were
computed long ago \cite{WWOhn,WWit,WZOhn,WZit,ZZOhn,ZZit} and have
recently been updated~\cite{CE}.  In Tables~\ref{tab:XStev} and
\ref{tab:XSlhc} we present the total cross sections for the various
processes at the Tevatron and LHC, for the MRST and CTEQ parton
distributions.  For the purpose of comparison we tabulated the results
for $\sigma^{\rm tot}$, the cross sections without any cuts applied,
but we also give $\sigma^{\rm cut}$, the cross sections with the
standard cuts defined in section~\ref{ComputationSection}.  At the
Tevatron, the $W^+Z$ and $W^-Z$ total cross sections are equal by CP
invariance.  The cross section values are for the scale $\mu =
(M_{V_1} + M_{V_2})/2$, where $M_{V_i}$ are the masses of the two
produced vector bosons.  Because the difference between the MRST and
CTEQ results is rather small, and given the fact that these
distributions will be updated regularly, we restrict ourselves in the
remainder of this paper to the MRST distribution. For
the same choices of input parameters and parton distributions, we
obtain perfect agreement with the total cross-sections tabulated in
Tables 1 through 4 of ref.~\cite{CE}.\footnote{Tables 1 and 2 of
ref.~\cite{CE} express their good agreement with
refs.~\cite{WWit,WZit,ZZit} for the older HMRSB parton distributions;
Tables 3 and 4 are for the Tevatron Run II and LHC with the MRST and
CTEQ(5) sets.}  The significant differences between the total cross
sections for the MRST distributions given in our Tables~\ref{tab:XStev} 
and \ref{tab:XSlhc} and those in ref.~\cite{CE} have their origin in
different input parameters, in particular $\sin^2\theta_W$. In the
case of the CTEQ set an additional difference is due to the
fact that in ref.~\cite{CE} the more recent CTEQ(5) parton
distributions have been used. 

\begin{table}
\begin{center}
\vskip0.2cm
\begin{tabular}{|c||c|c|c|c|c|c|} \hline
 & \multicolumn{2}{c|}{$ZZ$} & \multicolumn{2}{c|}{$W^+W^-$} 
 & \multicolumn{2}{c|}{$W^-Z$} \\ \cline{2-7} 
 & LO & NLO & LO & NLO & LO & NLO \\ \hline  \hline
$\sigma^{\rm tot}$(MRST) 
  & 1.13 & 1.44 & 9.52 & 12.4 & 1.37 & 1.84 \\ \hline
$\sigma^{\rm tot}$(CTEQ) 
  & 1.16 & 1.47 & 9.89 & 12.8 & 1.38 & 1.86 \\ \hline
$\sigma^{\rm cut}$(MRST) 
  & 0.352 & 0.446 & 3.17 & 4.22 & 0.377 & 0.506 \\ \hline
$\sigma^{\rm cut}$(CTEQ) 
  & 0.362 & 0.457 & 3.31 & 4.40 & 0.385 & 0.520 \\ \hline \hline
\end{tabular}
\end{center}
\caption[dummy]{\small Cross sections in ${\rm pb}$ for $p\bar{p}$ 
collisions at $\sqrt{s} = 2$ TeV. The statistical errors are $\pm 1$ 
within the last digit. 
\label{tab:XStev}}
\end{table}

\begin{table}
\begin{center}
\vskip0.2cm
\begin{tabular}{|c||c|c|c|c|c|c|c|c|} \hline
 & \multicolumn{2}{c|}{$ZZ$} & \multicolumn{2}{c|}{$W^+W^-$} 
 & \multicolumn{2}{c|}{$W^-Z$} & \multicolumn{2}{c|}{$W^+Z$} \\ \cline{2-9} 
 & LO & NLO & LO & NLO & LO & NLO & LO & NLO \\ \hline  \hline
$\sigma^{\rm tot}$(MRST) 
  & 11.4 & 15.2 & 77.9 & 115 & 11.0 & 19.0 & 17.6 & 30.1 \\ \hline
$\sigma^{\rm tot}$(CTEQ) 
  & 11.8 & 15.8 & 81.3 & 120 & 11.4 & 19.6 & 18.6 & 31.9 \\ \hline
$\sigma^{\rm cut}$(MRST) 
  & 3.95 & 5.31 & 24.6 & 40.4 & 3.41 & 6.44 & 5.08 & 9.38 \\ \hline
$\sigma^{\rm cut}$(CTEQ) 
  & 4.09 & 5.51 & 25.6 & 42.0 & 3.59 & 6.72 & 5.32 & 9.83 \\ \hline \hline
\end{tabular}
\end{center}
\caption[dummy]{\small Cross sections in ${\rm pb}$ for $p p$ collisions at
$\sqrt{s} = 14$ TeV. The statistical errors are $\pm 1$ within the last
digit.  \label{tab:XSlhc}}
\end{table}

The one-loop corrections to the total cross sections are of the order
of 50\% of the leading-order term. However, as we will see below, the
corrections can be much larger for large $p_T$ or invariant mass of
the vector bosons, particularly at the LHC. This is related to the
fact that at next-to-leading order the sub-processes $q g \to V_1 V_2 q$
have to be taken into account. These sub-processes generally dominate the
tail of the $p_T$-distribution (see e.g. ref.~\cite{WWit}).  Therefore,
a scale choice like 
\beq 
\mu^2 = \mu_{\rm st}^2 \equiv 
\frac{1}{2}(p_T^2(V_1)+p_T^2(V_2)+M_{V_1}^2+M_{V_2}^2)
\label{must}
\eeq 
seems to be more appropriate.  The difference between the two
different scale choices is very small for the total cross section, since
it is dominated by low $p_T$ vector bosons. However, for more exclusive
quantities the differences can be substantial. It is therefore necessary
to investigate the theoretical uncertainty related to the scale dependence
in some detail.

\begin{figure}
\centerline{ \epsfig{figure=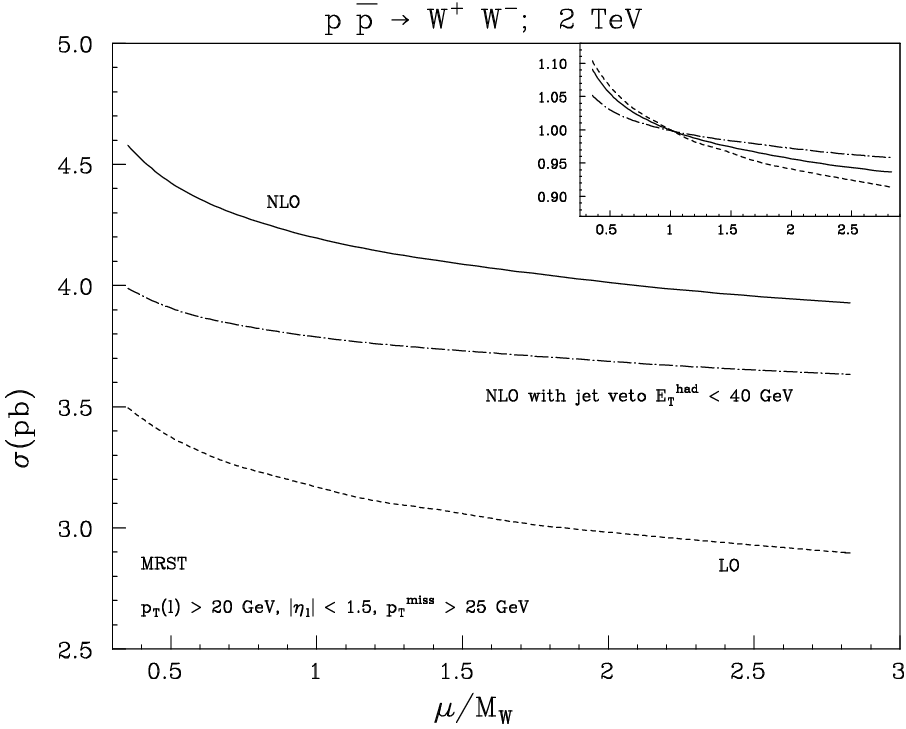,width=0.64\textwidth,clip=} }
   \ccaption{}{ \label{fig:XStev} Scale dependence of $\sigma^{\rm
   cut}$, the cross section for $W$-pair production at the
   Tevatron with standard cuts. The scale is given in units of
   $M_W$. We show the LO, NLO and NLO with jet veto curves. The inset
   shows the three curves normalized to 1 at $\mu = M_W$.  }
\end{figure}

To start with, in Fig.~\ref{fig:XStev} we consider the scale
dependence of the cross section for $W$-pair production at the
Tevatron.  We apply our standard cuts and vary the scale around $\mu =
M_W$.  The leading-order scale dependence is entirely due to the
decrease of quark distribution functions $q(x)$ with increasing
factorization scale for moderate $x$.  The NLO result does indeed have
a reduced scale dependence. We also show the NLO result with an
additional cut on the transverse hadronic energy, $E_T^{\rm had} <
40$~GeV, implemented at the parton level. In the remainder of this
paper we refer to this additional cut as jet veto, even though it does
not exactly correspond to a jet veto applied by experimentalists.
This additional cut reduces the scale dependence further.  These
results, together with the modest one-loop corrections we find below
for kinematic distributions at the Tevatron, lead to a rather
satisfactory description of $W$-pair production at this collider.

The situation is somewhat more delicate for the LHC. The one-loop
corrections can be huge in the tails of the distributions. We therefore
investigate the scale dependence in a more detailed way.  We again
consider the cross section with and without the same jet veto,
$E_T^{\rm had} < 40$~GeV.  We also consider the cross section
with an additional pair of cuts on the transverse momenta of the charged
leptons.  We require the larger of the two transverse momenta of the
charged leptons, $p_T^{\rm max}$, to be bigger than 200~GeV and the
smaller, $p_T^{\rm min}$, to be bigger than 100~GeV.  As discussed above,
it is therefore more appropriate to vary the scale around $\mu$ as given
in \eqn{must} instead of $\mu=M_W$.

\begin{figure}
\centerline{
   \epsfig{figure=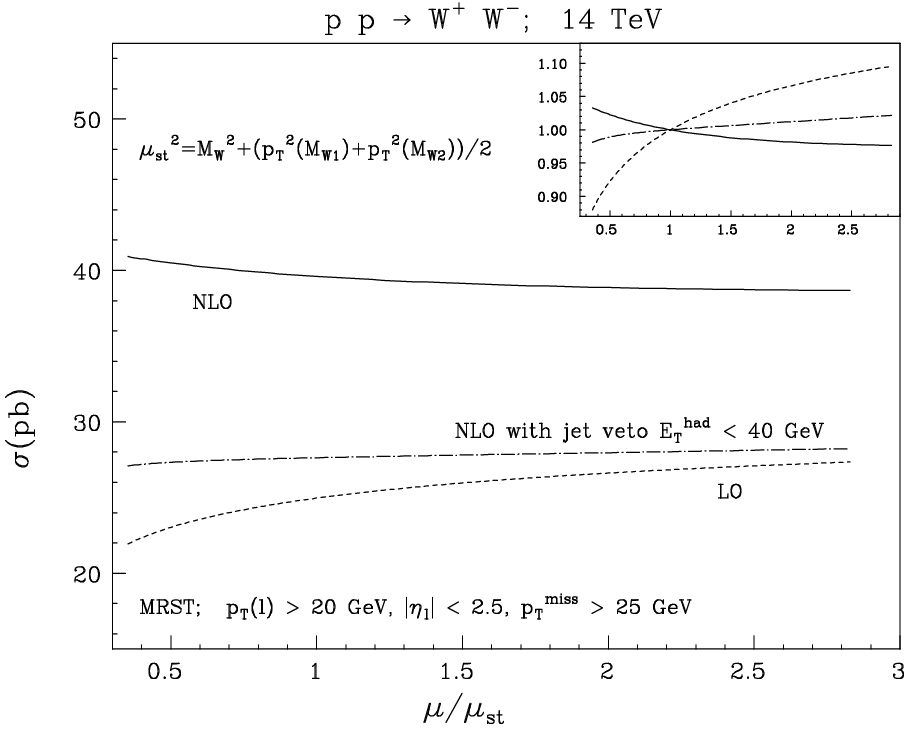,width=0.48\textwidth,clip=}
   \hfill
   \epsfig{figure=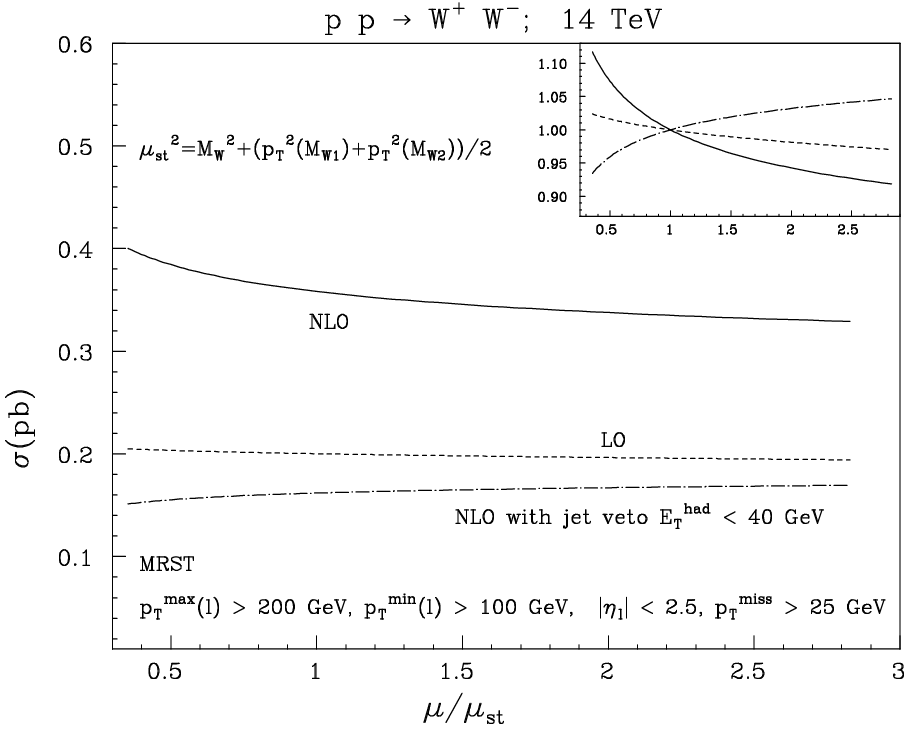,width=0.48\textwidth,clip=} }
\ccaption{}{ \label{fig:XSlhc} Scale dependence of $\sigma^{\rm cut}$,
   the cross section for $W$-pair production at the LHC with
   standard cuts. The scale is given in units of $\mu_{\rm st}$ as
   defined in \eqn{must}. We show the LO, NLO and NLO with jet veto
   curves without additional cuts (left) and with an additional cut
   $p_T^{\rm max}(\ell)>200$~GeV and $p_T^{\rm min}(\ell)>100$~GeV
   (right). The insets show the  curves normalized to 1 at
   $\mu=\mu_{\rm st}$.  }
\end{figure}                                                              

The purpose of these additional cuts is to investigate the scale
dependence of the cross section in the region where larger
corrections are expected. Indeed, the one-loop correction to the LHC
cross section for $\mu=\mu_{\rm st}$ increases from 60\% to 80\%
as the additional set of cuts is applied.  Before applying the 
additional cuts, the situation is very similar to the Tevatron. 
The scale dependence at LO is reduced at NLO, and is reduced even 
further if the jet veto is applied.  For the high $p_T$ case this 
is not quite true. The leading-order result is surprisingly stable 
under scale variations. This feature is somewhat artificial, however; 
we have checked that it does not hold if the additional cuts are 
changed to e.g. $p_T^{\rm max}(\ell)>400$~GeV and 
$p_T^{\rm min}(\ell)>200$~GeV. In this case,
the leading-order cross section decreases with increasing scale.
With the cuts $p_T^{\rm max}(\ell)>200$~GeV and $p_T^{\rm min}(\ell)>
100$~GeV we just happen to be close to the transition from a rising to
a falling leading-order cross section, a transition which is associated 
with the different behavior under evolution of the quark distribution 
functions at small $x$ vs. moderate $x$.

The reduction of the scale dependence of the next-to-leading order
results when a jet veto is applied seems to be quite general though. 
This situation is a bit paradoxical because the cross section 
with a jet veto is less inclusive and, therefore, expected to be more 
sensitive to large logarithms created by incomplete cancellation of the
infrared singularities.  On the other hand, any subprocess appearing 
at NLO in $pp\to WW$ produces additional hadronic energy in the final 
state. Thus, a cut on $E_{\rm had}$ naturally suppresses the one-loop 
corrections and tends to stabilize the perturbative expansion. 
This effect competes against the stronger sensitivity to large
logarithms.  Apparently, for the jet veto we applied, 
$E_T^{\rm had} < 40$~GeV, the subprocess-suppression effect still 
dominates.


\subsection{$W^+ W^-$}

In this subsection we present some results for kinematic distributions 
for the processes 
$p p\to W^-W^+\to \ell^- \bar\nu \ell^{\prime +} \nu^\prime$
and $p \bar{p}\to W^-W^+\to \ell^- \bar\nu \ell^{\prime +} \nu^\prime$.
Similar studies have been carried out earlier (see e.g.
ref.~\cite{OhnSpin}). Throughout, we apply our standard cuts as
defined in section~\ref{ComputationSection}. As an illustration, we
have chosen eight variables, four $p_T$-like quantities and four
angular distributions. They all are defined in terms of observable
momenta. The $p_T$-like distributions are defined as follows:
\begin{description}
\item{$p_T(\ell^-)$:  transverse momentum of the negatively charged lepton.}
\item{$M_{\ell\ell}$: invariant mass of the lepton pair.}
\item{$p_T^{\rm miss}$: missing transverse momentum, 
     $\sqrt{(\vec{p}_T(\ell^-) + \vec{p}_T(\ell^{\prime +}) +
      \vec{p}_T({\rm jet}))^2}$.} 
\item{$p_T^{\rm max}$: maximal transverse momentum of the two charged
      leptons, $\max\{p_T(\ell^-), p_T(\ell^{\prime +})\}$. }
\end{description}
In the case of the Tevatron, the $p_T(\ell^-)$ and
$p_T(\ell^{\prime +})$ distributions are equal.   However, this is not 
true for the LHC.

The four angular distributions we considered are defined as follows:
\begin{description}
\item{$\eta(\ell^-)$:  rapidity of the negatively charged lepton. }
\item{$\Delta\eta(\ell)$: rapidity difference between the leptons,
    $\eta(\ell^-)-\eta(\ell^{\prime +})$. }
\item{$\cos\theta_{\ell \ell}$: angle between the leptons,
   $\cos(\angle[\vec{p}(\ell^-),\vec{p}(\ell^{\prime +})])$. }
\item{$\cos\phi_{\ell \ell}$: transverse angle between the leptons,
    $\cos(\angle[\vec{p}_T(\ell^-),\vec{p}_T(\ell^{\prime +})])$. } 
\end{description}
For massless leptons, the true rapidity 
$y(\ell) = {1\over2} 
\ln\Bigl( { E(\ell)+p_L(\ell) \over E(\ell)-p_L(\ell) } \Bigr)$
is equal to the pseudorapidity $\eta(\ell) = -\ln(\tan(\theta/2))$,
so we refer to both as rapidity.
For the Tevatron, we have to specify the proton direction:
it corresponds to $\theta = 0$. 

At the Tevatron, the rapidity distribution for $\ell^{\prime +}$ can
be obtained trivially from that for $\ell^-$ by changing
the sign of $\eta$.  For the LHC, there is no such simple relation
between the two distributions.   Note also that the
$\Delta\eta(\ell)$ distribution is not symmetric around $\eta=0$ for
the Tevatron; it is symmetric for the LHC. The $\cos\theta_{\ell\ell}$ 
observable has been investigated in ref.~\cite{DD} in the
context of Higgs boson detection in the intermediate Higgs mass
range $m_H=155$--180~GeV. It is therefore particularly interesting to
consider the effect of the QCD corrections to this observable.
 
\begin{figure}
\centerline{ \epsfig{figure=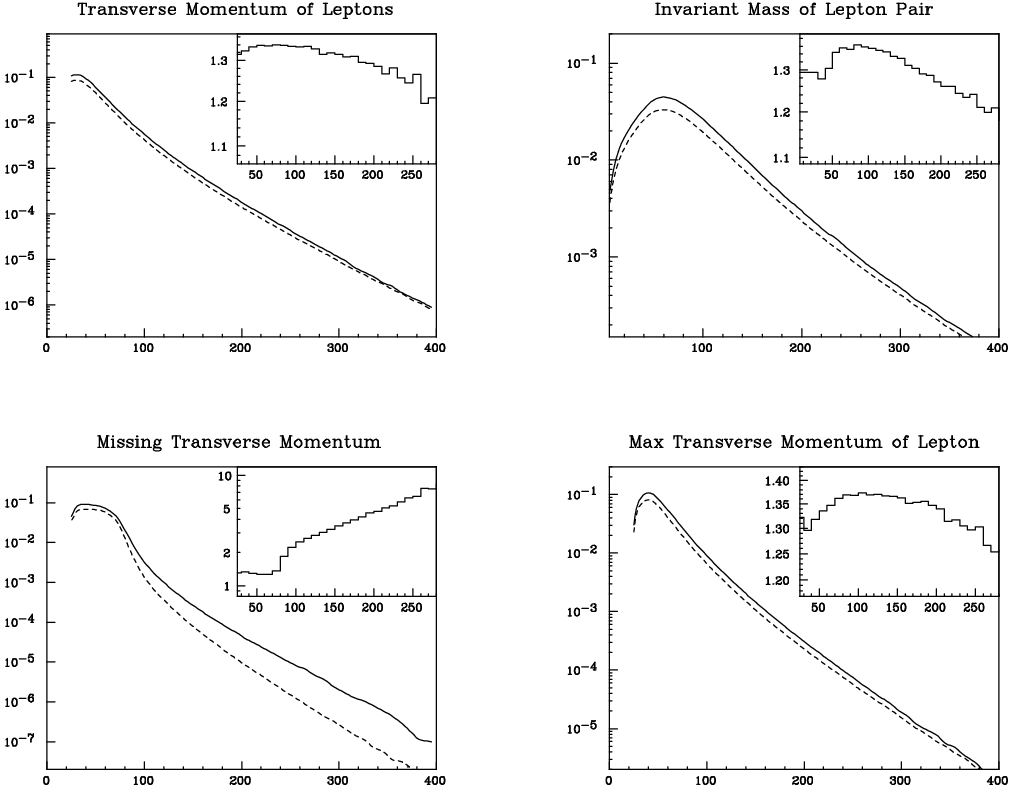,width=0.94\textwidth,clip=} }
   \ccaption{}{Differential cross sections in pb/GeV for $W^-W^+$
   production at the Tevatron for the $p_T$-like variables
   $p_T(\ell^-)$, $M_{\ell\ell}$, $p_T^{\rm miss}$ and $p_T^{\rm max}$
   defined in the text at LO (dashed curves) and NLO (solid curves),
   with $\mu=M_W$. Standard cuts have been applied and the branching
   ratios for the leptonic decay of the vector bosons are not
   included. The insets show the ratio $d\sigma^{NLO}/d\sigma^{LO}$.
   The units on the horizontal axes are GeV.  \label{fig:D1tev} }
\end{figure}                                                              

\begin{figure}
\centerline{ \epsfig{figure=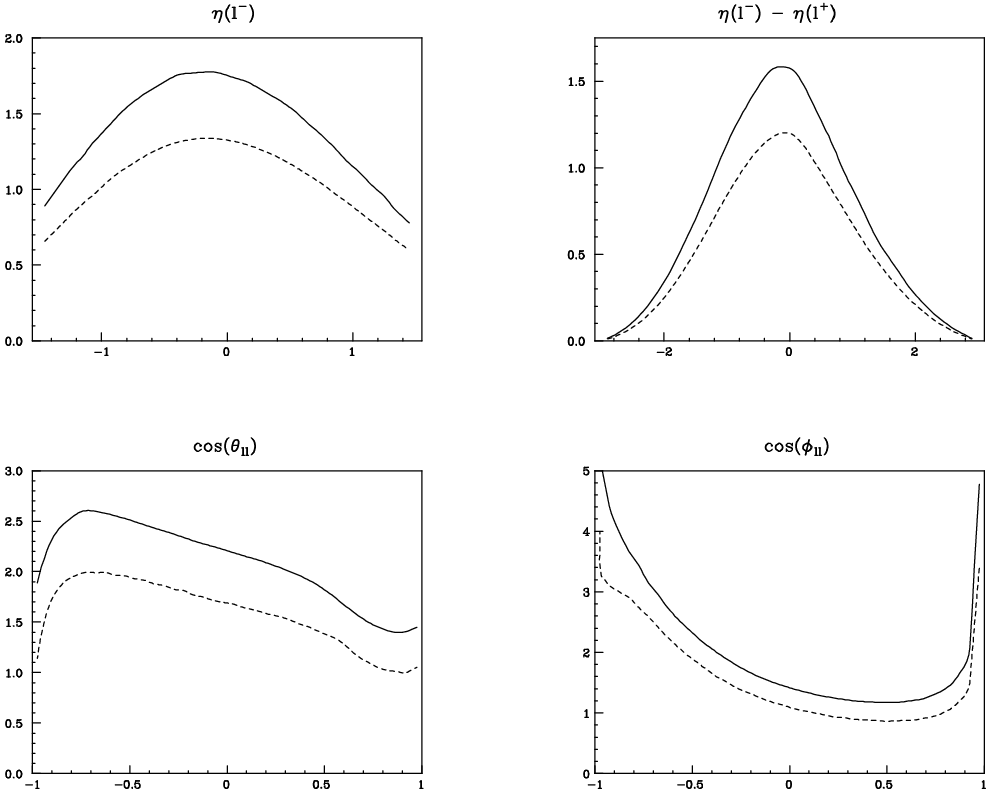,width=0.94\textwidth,clip=} }
   \ccaption{}{Differential cross sections in pb for the Tevatron for
   the $W^-W^+$ angular variables defined in the text at LO (dashed
   curves) and NLO (solid curves), with $\mu=M_W$. Standard cuts have
   been applied and the leptonic branching ratios are not
   included. \label{fig:D2tev} }
\end{figure}                                                              

In Figs.~\ref{fig:D1tev} and \ref{fig:D2tev} we show the
distributions for the Tevatron, evaluated at $\mu=M_W$ and with our
standard cuts applied.  Recall that the branching ratios for the
leptonic decay of the vector bosons are not included. The perturbative
expansion for angular-type distributions is generally better behaved
than that for steeply falling $p_T$-like distributions. The latter 
often suffer from large NLO corrections in the tails of the distribution. 
The insets of Fig.~\ref{fig:D1tev} show the ratio 
$d\sigma^{NLO}/d\sigma^{LO}$.  They indicate that for the Tevatron, the 
NLO corrections are not too large, except for the $p_T^{\rm miss}$ 
distribution.  

The large corrections to the $p_T^{\rm miss}$ distribution reflect a
suppression of the Born-level cross section when $p_T^{\rm miss}$ is
large~\cite{BHOWW}:  The only way to get a large $p_T^{\rm miss}$ at Born
level is to have a large $WW$ invariant mass.  Then the two neutrinos are
almost back-to-back in the $WW$ rest frame.  Also, for large invariant
mass the Standard Model $WW$ helicity amplitudes are dominantly those
where the $W^+$ and $W^-$ have opposite helicities.  The $V-A$ decay of
the $W$ bosons then implies that two neutrinos tend to carry roughly the
same fraction of the $W$ momentum, i.e. they have cancelling transverse
momenta.  NLO QCD corrections have a big effect at large $p_T^{\rm miss}$
because they allow a recoiling final-state parton to spoil the $p_T$
balance of the two neutrinos.  The presence of anomalous couplings can
also have a big effect in this region, by relaxing the helicity
anti-correlation of the two $W$ bosons~\cite{BHOWW}.

\begin{figure}
\centerline{ \epsfig{figure=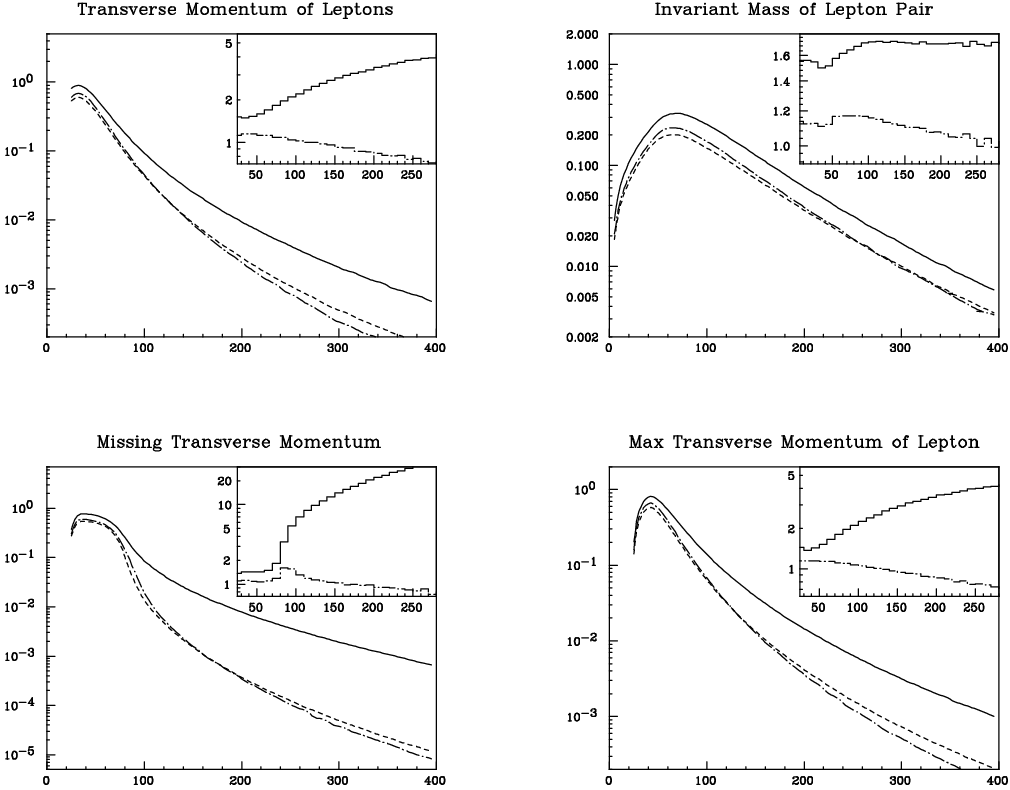,width=0.94\textwidth,clip=} }
   \ccaption{}{Differential cross sections in pb/GeV for $W^-W^+$
   production at the LHC, for the $p_T$-like variables defined in the
   text at LO (dashed curves) and NLO (solid curves), with
   $\mu=M_W$. Standard cuts have been applied and the leptonic
   branching ratios are not included. Also shown (as dot-dashed lines)
   are the NLO curves with a jet veto $E_T^{\rm had} < 40$~GeV. The
   insets show the ratio $d\sigma^{NLO}/d\sigma^{LO}$.  The units on
   the horizontal axes are GeV.  \label{fig:D1lhc} }
\end{figure}

\begin{figure}
\centerline{ \epsfig{figure=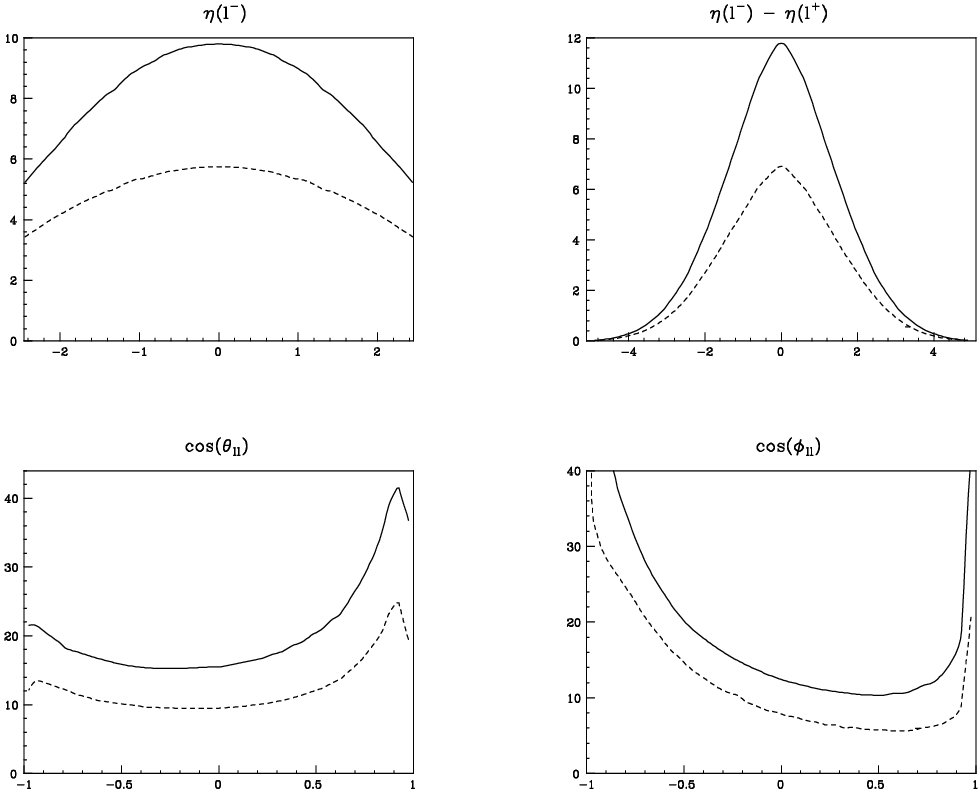,width=0.94\textwidth,clip=} }
   \ccaption{}{Differential cross sections in pb for the LHC for the
   $W^-W^+$ angular variables defined in the text at LO (dashed
   curves) and NLO (solid curves), with $\mu=M_W$. Standard cuts have
   been applied and the leptonic branching ratios are not
   included. \label{fig:D2lhc} }
\end{figure}

In Figs.~\ref{fig:D1lhc} and \ref{fig:D2lhc}, the same two sets of 
distributions are displayed for the LHC, again for $\mu = M_W$. 
The NLO corrections to the angular distributions are again modest.
The $p_T$-like distributions, however, can have much larger
corrections, particularly in their tails, where the NLO result can 
easily exceed the leading order result by a factor of 5. 
The $p_T^{\rm miss}$ distribution has the largest corrections of all, 
for the reason mentioned above, and the factor can be 20 or more.
As mentioned before, the huge corrections in the tail of the 
generic $p_T$-like distributions have to do with
the subprocess $qg\to WWq$ which dominates in this kinematical
region.  It may be argued that a scale choice as in \eqn{must} is
mandatory in this case. However, we have checked that such a scale
choice does not lead to a substantial improvement. 

The problem is that in a NLO calculation for $pp\to WW$ the partonic
process $\bar{q}q\to WW$ is included at NLO, but the $qg\to WWq$
subprocess is included only at LO. Therefore, in a kinematical region
where the latter dominates, the calculation presented in this paper is
effectively only a leading order calculation. The only fully satisfactory
way to improve the theoretical prediction in such cases is to include
the one-loop corrections to the subprocess with a $qg$ in the initial
state. These $\Ord(\as^2)$ corrections correspond to a NNLO contribution 
to $pp\to WW$. At the same order in $\as$, there are two new subprocesses: 
$gg\to WW$ at one loop, and $gg\to WW q\bar{q}$ at tree level. 
Due to the large gluon density at small $x$, these contributions are also
expected to be important for LHC energies, and would have to be included if
a reliable prediction for the tails of the $p_T$-like distributions were 
required. 

As discussed above, a way to suppress the partonic processes with
gluons in the initial state is to impose a jet veto. The effect of the
cut $E_T^{\rm had} > 40$~GeV on the $p_T$-like distributions is even
more dramatic than its effect on the total cross section. This can be
seen in Fig.~\ref{fig:D1lhc}, where we also show the NLO curves with
the jet veto (dot-dashed lines). Indeed, the one-loop corrections are
very small throughout.  We conclude that if a reliable theoretical
prediction is desired for the tail of a $p_T$-like distribution, and
only a next-to-leading order program is available, then a jet veto is
unavoidable.

\subsection{$W^\pm Z$}

In this subsection we study $WZ$ production, followed by leptonic decay of
each boson.   In particular, we examine the effects 
of QCD corrections on the approximate radiation zero at 
$\cos\theta = (g^-_1 + g^-_2)/(g^-_1 - g^-_2)$~\cite{BHOWZZero}.
Since the precise flight direction of the $W$ boson is not known, due to 
the uncertainty in the longitudinal momentum carried by the neutrino, we 
simply choose to plot a distribution in the (true) rapidity difference
between the $Z$ boson and the charged lepton coming from the
decay of the $W$, $\Delta y_{Z\ell} \equiv y_Z - y_\ell$.
This quantity is similar to the rapidity difference 
$\Delta y_{WZ} \equiv | y_W - y_Z |$ studied in ref.~\cite{WZit}, but
uses only the observable charged-lepton variables.  It is the direct
analog of the variable $y_\gamma-y_{\ell^+}$ considered in ref.~\cite{BEL}
for the case of $W\gamma$ production.  (It is possible 
to determine $\cos\theta$ in the $W\gamma$ or $WZ$ rest frame, 
by solving for the neutrino longitudinal momentum using the 
$W$ mass as a constraint, up to a two-fold discrete ambiguity for each 
event~\cite{nulong}.  However, ref.~\cite{BEL} found that the
ambiguity degrades the radiation zero --- at least if each solution is 
given a weight of 50\% --- so that the rapidity difference 
$y_\gamma-y_{\ell^+}$ is more discriminating than $\cos\theta$.)

\begin{figure}
\centerline{ \epsfig{figure=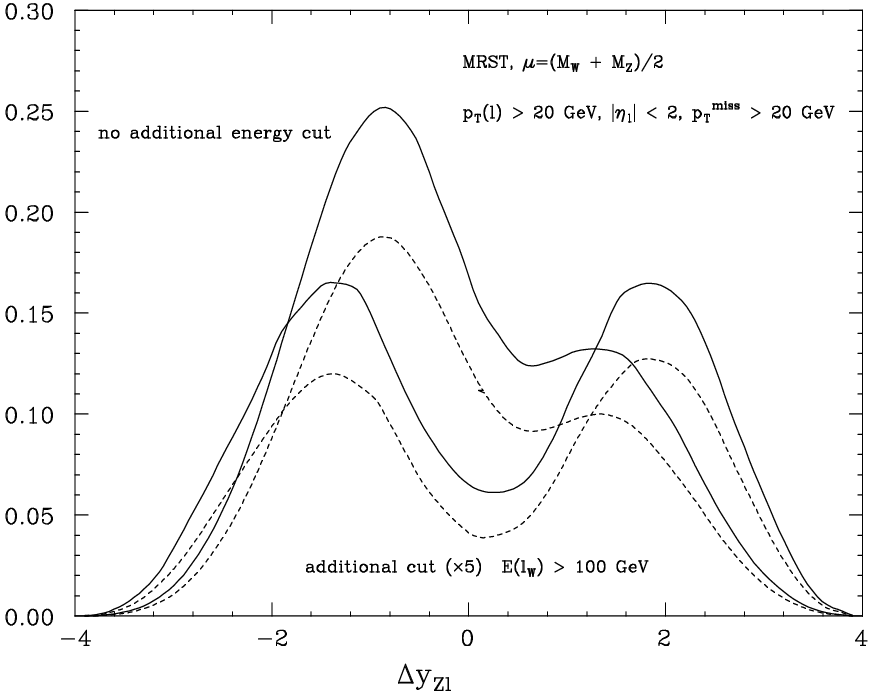,width=0.64\textwidth,clip=} }
   \ccaption{}{ \label{fig:RadZFig} For $WZ$ production at Run II,
   followed by leptonic decays of both the $W$ and $Z$ bosons, we plot
   the distribution, in picobarns, in the rapidity difference between
   the $Z$ and the charged lepton $\ell$ from the decay of the $W$,
   $\Delta y_{Z\ell} \equiv y_Z - y_\ell$.  Leptonic branching ratios
   are not included and the scale has been set to $\mu =
   (M_W+M_Z)/2$. The basic cuts used are $p_T(\ell) > 20$ GeV and
   $|\eta(\ell)| < 2$ for all three charged leptons, and a missing
   transverse momentum cut of $p_T^{\rm miss} > 20$ GeV.  We plot the
   $\Delta y_{Z\ell}$ distribution with these cuts, and also after
   imposing an additional cut on the $W$ decay lepton, $E(\ell) >
   100$~GeV; the latter curves have been scaled up by a factor of 5.
   The dashed curves are Born-level results; the solid curves include
   the $\Ord(\alpha_s)$ corrections.  }
\end{figure}

As one can see from~\fig{RadZFig}, there is a residual dip in the
$\Delta y_{Z\ell}$ distribution, even at order $\alpha_s$. This dip
can easily be enhanced by requiring a minimal energy for the decay
lepton from the $W$. In \fig{RadZFig}, we have chosen
$E(\ell)>100$~GeV.  Note that these curves are scaled up by a factor
5.  Unfortunately, only a few tens of $WZ \to$ leptons events are
expected at Run II of the Tevatron, so the observation of such a
dip will be rather difficult prior to the LHC.

\section{Anomalous $W^+W^-Z$ and $W^+W^-\gamma$ Couplings}
\label{AnomalousSection}

\subsection{Triple Gauge Boson Vertices}

New physics may modify the self-interactions of vector bosons, in
particular the triple gauge boson vertices.  If the new physics occurs 
at an energy scale well above that being probed experimentally, it
can be integrated out, and the result expressed as a set of 
anomalous (non-Standard Model) interaction vertices.  

Here we consider anomalous $W^+W^-Z$ and $W^+W^-\gamma$ trilinear
couplings, and their effects on the hadronic production of $WW$ and $WZ$
pairs up through order $\alpha_s$.  The most general set of Lagrangian 
terms for $WWV$, $V \in \{Z,\gamma\}$, that conserves $C$ and $P$ 
separately, is (see e.g. \cite{Hagi,EllisonWudka}) 
\bea
{\cal L}/g_{WWV} = i g_1^V (W_{\mu \nu}^{\ast} W^\mu V^\nu - W_{\mu \nu}
W^{\ast \ \mu} V^\nu) 
+ i \kappa^V W_{\mu}^{\ast} W_\nu V^{\mu\nu} + i
\frac{\lambda^V}{M_W^2} W_{\rho \mu}^{\ast} W^\mu_{~\nu} V^{\nu \rho} \,,
\label{AnomL}
\eea
where $X_{\mu \nu} \equiv \partial_\mu X_\nu - \partial_\nu X_\mu $ and the
overall coupling constants $g_{WWV}$ are given by $g_{WW\gamma} = - e$ and
$g_{WWZ} = -e\, \cot \theta_W$ respectively, with $\theta_W$ the weak
mixing angle.  The Standard Model triple gauge boson vertices are
recovered by letting $g_1^V \to 1, \kappa^V \to 1$ and $\lambda^V \to 0$. 
The coupling factors can be written in terms of their deviation from Standard
Model values:  $g_1^V = 1 + \Delta g_1^V$ and 
$\kappa^V = 1 + \Delta \kappa^V$.   Electromagnetic gauge invariance 
requires $g_1^\gamma = 1$, or $\Delta g_1^\gamma = 0$.  Sometimes
other constraints are imposed on the couplings.  For example, if one
requires the existence of an effective Lagrangian with $SU(2)\times U(1)$ 
invariance, and neglects operators with dimension eight or higher,
then the number of independent coefficients in \eqn{AnomL} is 
reduced from five to three,
\bea
 \Delta g_1^Z = {\alpha_{W\phi} \over \cos^2\theta_W}\,,&&
\hskip 2 cm
 \lambda^\gamma = \lambda^Z = \alpha_W, \nn \\
 \Delta\kappa^\gamma = \alpha_{W\phi} + \alpha_{B\phi}\,,&&
\hskip 2 cm
 \Delta\kappa^Z = \alpha_{W\phi} 
 - {\sin^2\theta_W \over \cos^2\theta_W} \, \alpha_{B\phi},
\eea
where $\alpha_W$, $\alpha_{W\phi}$ and $\alpha_{B\phi}$ are coefficients
of the dimension-six operators in this effective 
Lagrangian~\cite{EllisonWudka}.
If one arbitrarily supposes that $\alpha_{W\phi} = \alpha_{B\phi}$,
one arrives at the so-called HISZ scenario~\cite{HISZ}, with only 
two independent couplings.

The momentum-space vertex $W^-_\alpha(q) W^+_\beta(\qb) V_\mu(p)$
(where $p + q + \qb = 0$) corresponding to \eqn{AnomL} can be written as
\bea
\Gamma^{\alpha \beta \mu}(q, \qb, p)/g_{WWV} &=& 
  \qb^\alpha g^{\beta \mu} 
    \biggl( g_1^V + \kappa^V + \lambda^V {q^2\over M_W^2} \biggr) 
 - q^\beta g^{\alpha \mu}
    \biggl( g_1^V + \kappa^V + \lambda^V {\qb^2\over M_W^2} \biggr) \nn \\ 
&& \hskip 1 cm
 + \bigl( \qb^\mu - q^\mu \bigr) 
 \Biggl[ - g^{\alpha \beta} \biggl( 
   g_1^V + {1\over2} p^2 \frac{\lambda^V}{M_W^2} \biggr) 
 +\frac{\lambda^V}{M_W^2} p^\alpha p^\beta \Biggr] \,.
\eea
%
%
%
Here we have used momentum conservation, and the fact that the terms 
$q^\alpha, \qb^\beta$ and $p^\mu$ can be neglected, but we have not 
imposed on-shell conditions on the vector bosons.  As it stands,
this vertex will eventually lead to a violation of unitarity.  
To avoid this, the deviations from the Standard Model, 
$\Delta g_1^V, \Delta \kappa^V$ and $\lambda^V$, have to be supplemented 
with form factors.  Since the form factors are supposed to be produced
by unknown physics, the form they should take is {\it a priori} somewhat 
arbitrary.  We choose a conventional dipole form factor, i.e.
\beq
\Delta g_1^V \to { \Delta g_1^V \over (1 + \hat{s}/\Lambda^2)^2 }\,, \qquad
\Delta \kappa^V \to { \Delta \kappa^V \over (1 + \hat{s}/\Lambda^2)^2 }\,,
\qquad
\lambda^V \to { \lambda^V \over (1 + \hat{s}/\Lambda^2)^2 } \,,
\label{FormFactor}
\eeq
where $\hat{s}$ is the invariant mass of the vector boson pair, and 
$\Lambda$ is in the TeV range.


\subsection{Tree, Virtual and Bremsstrahlung Amplitudes for $WW$}

Replacing the Standard Model vector-boson-vertex by the more general
vertex given above results in the some modifications of the primitive
amplitudes presented in ref.~\cite{DKS}. We use the same notation in this
paper and refer the reader to ref.~\cite{DKS} for more details.  The
box-parent primitive amplitudes are not affected by changes in the trilinear
vector-boson-vertex.  The change in the triangle-parent primitive amplitude
can be obtained by simply computing the one tree-level diagram with
the new vertex.  Since the vertex is no longer symmetric in the
exchange $W\lr V$, we get slightly different results for the different
final states.  For the $WW$ final state, the new tree amplitude, which 
replaces $A^{{\rm tree},b}$ in eq.~(2.9) of ref.~\cite{DKS}, is
\bea
A^{\tree,B} &=& \frac{i}{ 2\,s_{12}\,s_{34}\,s_{56} } 
\Bigg[(g_1^V + \kappa^V + \lambda^V) 
   \bigl( \spa1.3\spb2.4\spab6.{(1+2)}.5 +
          \spa1.6\spb2.5\spab3.{(5+6)}.4 \bigr) \nn \\
&+& \spab1.{(3+4)}.2 \biggl( 2 g_1^V \, \spa3.6\spb4.5
     + \frac{\lambda^V}{M_W^2} \, \spab3.{(1+2)}.5 \spab6.{(1+2)}.4 \biggr) 
\Bigg] \,.
\label{AnomTreeWW} 
\eea
For the limit $g_1^V \to 1, \kappa^V \to 1, \lambda^V \to 0$, we recover 
the Standard Model result,
\bea
A^{\tree,b}  &=&  { i \over s_{12}\,s_{34}\,s_{56} }
\Bigl[ -\, \spa3.6\spb4.5 \spab1.{(5+6)}.2
       + \spa1.3\spb2.4 \spab6.{(1+2)}.5 \nn\ \\
&& \qquad \qquad\ \   +\, \spa1.6\spb2.5 \spab3.{(5+6)}.4 \Bigr] \nn\ \\
&=&
{ i \over s_{12}\,s_{34}\,s_{56} }
\Bigl[ \spa1.3\spb2.5 \spab6.{(2+5)}.4
     + \spb2.4\spa1.6 \spab3.{(1+6)}.5 \Bigr] \,.
\label{treeb}
\eea

With the above results we also get immediately the one-loop primitive
amplitude for anomalous couplings. Assuming the usual decomposition into
finite and divergent pieces, the finite pieces are vanishing, as in
the Standard Model case, and the divergent pieces are still given by
$ \cg A^{\tree,B} V $ where
\beq
V = 
  - {1\over \eps^2} \left( {\mu^2 \over -s_{12}}\right) ^\eps 
  - {3\over 2 \eps} \left( {\mu^2 \over -s_{12}}\right) ^\eps 
  - {7\over 2} \,. 
\,
\label{Vab}
\eeq


The result for the bremsstrahlung diagrams with an additional positive 
helicity gluon radiated off the quark line is given by
\bea
A_7^{\tree,B} &=& 
\frac{i}{2\,\spa1.7\spa7.2 s_{34}\,s_{56}\,t_{127}} \nn \\
&&\hskip -0.4 cm
  \times  \Biggl[  (g_1^V + \kappa^V + \lambda^V) 
      \bigl( \spa1.3 \spab1.{(2+7)}.4 \spab6.{(3+4)}.5 -
             \spa1.6 \spab1.{(2+7)}.5 \spab3.{(5+6)}.4 \bigr) \nn \\
&& \hskip 0.4 cm   + \, \spaa1.{(3+4)(2+7)}.1 
   \biggl( 2 g_1^V \, \spa3.6\spb4.5
   + \frac{\lambda^V}{M_W^2} \, \spab3.{(4+6)}.5 \spab6.{(3+5)}.4 \biggr) 
\Biggr] \,,
\label{AnomBremWW}
\eea
replacing $A_7^{{\rm tree},b}$ in eq.~(2.22) of ref.~\cite{DKS}.
The result for a negative helicity gluon can be obtained by the usual flip
operation~\cite{DKS}.

We also have to modify the prescription in ref.~\cite{DKS} for 
dressing the above primitive amplitudes with electroweak
couplings. Since $g_1^V, \kappa^V $ and $\lambda^V$ are relative
couplings, i.e. the overall coupling $g_{WWV}$ has not been changed, the
dressing with electroweak factors is almost identical to the Standard
Model case.  The only subtlety is that both $Z$ and $\gamma$ appear as
intermediate states in $WW$ production.  In the coefficient functions
\bea
C_{L,\{ {u\atop d} \}} &=& \pm 2 Q \sstw 
+ {s_{12} ( 1 \mp 2 Q \sstw ) \over s_{12} - M_Z^2} \,,
\label{CleftWWZ} \\
C_{R,\{ {u\atop d} \}} &=& \pm 2 Q \sstw 
\mp  2 Q \sstw \frac{s_{12}}{s_{12} - M_Z^2} \nn \, ,
\eea
defined in ref.~\cite{DKS}, the first term ($\pm 2 Q \sstw$) is from
the intermediate $\gamma$, while the second term is from the
intermediate $Z$.  Correspondingly, we should set $V=\gamma$ ($V=Z$)
in eqs.~(\ref{AnomTreeWW}) and (\ref{AnomBremWW}), when they are
dressed with the first (second) term in $C_{\{ {L\atop R} \},\{ {u\atop
d} \}}$.  Otherwise, all the prefactors remain the same and only the
`new' primitive amplitudes have to be plugged in.


\subsection{Tree, Virtual and Bremsstrahlung Amplitudes for $WZ$}

For the $WZ$ final state, the new tree primitive amplitude is
\bea
A^{\tree,B} &=& \frac{-i}{ 2\,s_{12}\,s_{34}\,s_{56} } 
\Bigg[ \Bigl( g_1^Z + \kappa^Z + \lambda^Z \frac{s_{12}}{M_W^2} \Bigr)
          \spa3.6 \spb4.5 \spab1.{(5+6)}.2  \nn \\
&& \hskip 2.3 cm
+ \, (g_1^Z + \kappa^Z + \lambda^Z) \spa1.6 \spb2.5 \spab3.{(1+2)}.4 \nn \\
&& \hskip 1 cm 
  + \spab6.{(3+4)}.5 \biggl( 2 g_1^Z \, \spa1.3\spb2.4
     + \frac{\lambda^Z}{M_W^2} \, \spab3.{(5+6)}.2 \spab1.{(5+6)}.4 \biggr) 
\Bigg] \,.
\label{AnomTreeWZ} 
\eea
The new bremsstrahlung primitive amplitude is
\bea
A_7^{\tree,B} &=& 
\frac{-i}{2\,\spa1.7\spa7.2 s_{34}\,s_{56}\,t_{127}} \nn \\
&&\times  \Biggl[  
  \Bigl( g_1^Z + \kappa^Z + \lambda^Z \frac{t_{127}}{M_W^2} \Bigr)
          \spa3.6 \spb4.5 \spaa1.{(5+6)(2+7)}.1  \label{AnomBremWZ} \\
&& \hskip 0.4 cm
  + \, (g_1^Z + \kappa^Z + \lambda^Z) 
         \spa1.6 \spab1.{(2+7)}.5 \spab3.{(5+6)}.4 \nn \\
&& \hskip 0.4 cm   + \, \spab6.{(3+4)}.5 
   \biggl( - 2 g_1^Z \, \spa1.3 \spab1.{(2+7)}.4  \nn \\
&& \hskip 3 cm
   + \frac{\lambda^Z}{M_W^2} \, \spaa3.{(5+6)(2+7)}.1 \spab1.{(5+6)}.4 
\biggr) \Biggr] \,. \nn
\eea
In this case, since there is only an intermediate $W$, the dressing with
electroweak coupling factors is indeed identical to the Standard Model case.


\subsection{Numerical Results}
\label{AnomResultsSubsection}

More systematic studies of the effects of anomalous couplings on hadronic
production of vector boson pairs have been carried out 
elsewhere~\cite{BZ,BHOWg,BHOZg,BHOWW,BHOWZ}.  Here we merely consider 
one sample distribution.

The effect of anomalous couplings is enhanced for gauge bosons that are
produced at large transverse momentum.  In order to exploit this feature,
the D0 collaboration considered a double-binned $E_T$ spectrum for the 
charged leptons coming from the decay of $W$ pairs~\cite{D0et}. 
Any deviation from the Standard Model should be more pronounced
in the high~$E_T$ bins.

\newcommand{\arr}[3]{$ \begin{array}{c} #1 \\ #2 \\ #3 \end{array}
 \  $}
\newcommand{\ten}[1]{10^{- #1}}
\newcommand{\pb}{{\rm pb}}
\begin{table}
\begin{center}
\vskip0.2cm
\begin{tabular}{|c||c|c|c|c|c|} \hline
 $E_T^{\rm max}$ $\backslash$ $E_T^{\rm min}$ & $20-38.1$ & $38.1-72.5$ &
$72.5-138$ & $138-263$  & $263-500$  \\ \hline\hline
$\begin{array}{c} 20 - 38.1 \\ \left[\pb\right] \end{array} $  
              & \arr{1.07}{1.07}{1.03} 1.10 
              & -- & -- & -- & -- \\ \hline
$\begin{array}{c} 38.1 - 72.5 \\ \left[\pb\right] \end{array} $ 
              & \arr{1.62}{1.62}{ 1.54} 1.61
              & \arr{0.77}{0.76}{0.74} 0.77
              & -- & -- & --  \\ \hline
$\begin{array}{c} 72.5 - 138\\ \left[\ten{1}\pb\right] \end{array} $ 
              & \arr{1.67}{1.60}{1.49} 1.76
              & \arr{3.30}{3.31}{3.15} 3.56
              & \arr{1.30}{1.32}{1.27} 1.50
              & -- & --   \\ \hline
$\begin{array}{c} 138 - 263\\ \left[\ten{2}\pb\right] \end{array} $ 
              & \arr{0.63}{0.57}{0.49} 2.43
              & \arr{1.27}{1.20}{1.09} 4.18
              & \arr{3.07}{3.16}{2.98} 6.90
              & \arr{1.08}{1.14}{1.09} 2.32
              & --   \\ \hline
$\begin{array}{c} 263 - 500 \\ \left[\ten{4}\pb\right] \end{array} $
              & \arr{0.7}{0.6}{0.5}  22 
              & \arr{1.2}{1.1}{0.9}  37 
              & \arr{2.6}{2.5}{2.2}  55 
              & \arr{9.0}{9.8}{9.0}  63 
              & \arr{2.3}{2.5}{2.4}  9.5 \\ \hline \hline
\end{tabular}
\end{center}
\caption[dummy]{\small Double-binned $E_T$ cross sections for $p\bar{p}\to
W^+W^-\to$ leptons at $\sqrt{s} = 2$~TeV. The three numbers in the left
column in each entry are the Standard Model results for the scales 
$\mu=\mu_{\rm st}\times\{1/2,1,2\}$, with $\mu_{\rm st}$ given in 
\eqn{must}. The fourth number in the entry is the cross section 
with anomalous couplings, as defined in the text. \label{tab:AC}}
\end{table}

We have computed a similar double-binned $E_T$ spectrum at NLO 
for Run II of the Tevatron.  As in ref.~\cite{D0et} we compute for 
each event the larger and smaller transverse energies of the two 
leptons, $E_T^{\rm max}$ and $E_T^{\rm min}$.   ($E_T$ is equivalent to
$p_T(\ell)$, of course.)  We impose our standard event cuts and then bin 
each $E_T$ into five bins with the following limits in GeV:
\beq
E_T = \{20,  38.1, 72.5, 138, 263, 500 \}.
\eeq
In order to get a feeling for the theoretical uncertainties, we
repeated the Standard Model computation for three scales, 
$\mu_{\rm st}/2$, $\mu_{\rm st}$ and $2\mu_{\rm st}$, where
$\mu_{\rm st}$ is given in \eqn{must}.
We computed the same NLO double-binned cross section
including the effects of anomalous couplings.  As an illustration we 
have chosen the HISZ scenario with 
$\alpha_W = \alpha_{W\phi} = \alpha_{B\phi} =
0.1$. This corresponds to the following values for the anomalous
couplings appearing in the Lagrangian in \eqn{AnomL}:
\beq
 \Delta g_1^\gamma = 0; \ \ \Delta g_1^Z = 0.13; \ \ 
 \lambda^\gamma = \lambda^Z = 0.1 ; \ \ 
 \Delta\kappa^\gamma = 0.2 ; \ \ 
 \Delta\kappa^Z = 0.07.
 \label{AnomCp}
\eeq 
The form factors have been chosen according to \eqn{FormFactor} with
$\Lambda=2$~TeV.  At present these values are still consistent with LEP2
bounds~\cite{LEP2anompresent}.  Table~\ref{tab:AC} presents the results.
As usual, the leptonic branching ratios are not included, and we use 
the MRST parton distributions.  The three numbers in the left column of 
each box give the Standard Model result for the three choices of $\mu$,
with $\mu$ increasing from top to bottom.  The fourth
number in each box is the result with the anomalous couplings
included and $\mu=\mu_{\rm st}$.  Units for the cross sections for
each row are given in square brackets.  As expected, the results for the
Standard Model and anomalous cross section are very similar for the
low $E_T$ bins. For the high $E_T$ bins, however, the differences are
large and certainly much bigger than the most conservative estimate of
the theoretical error.  

\begin{figure}
\centerline{
   \epsfig{figure=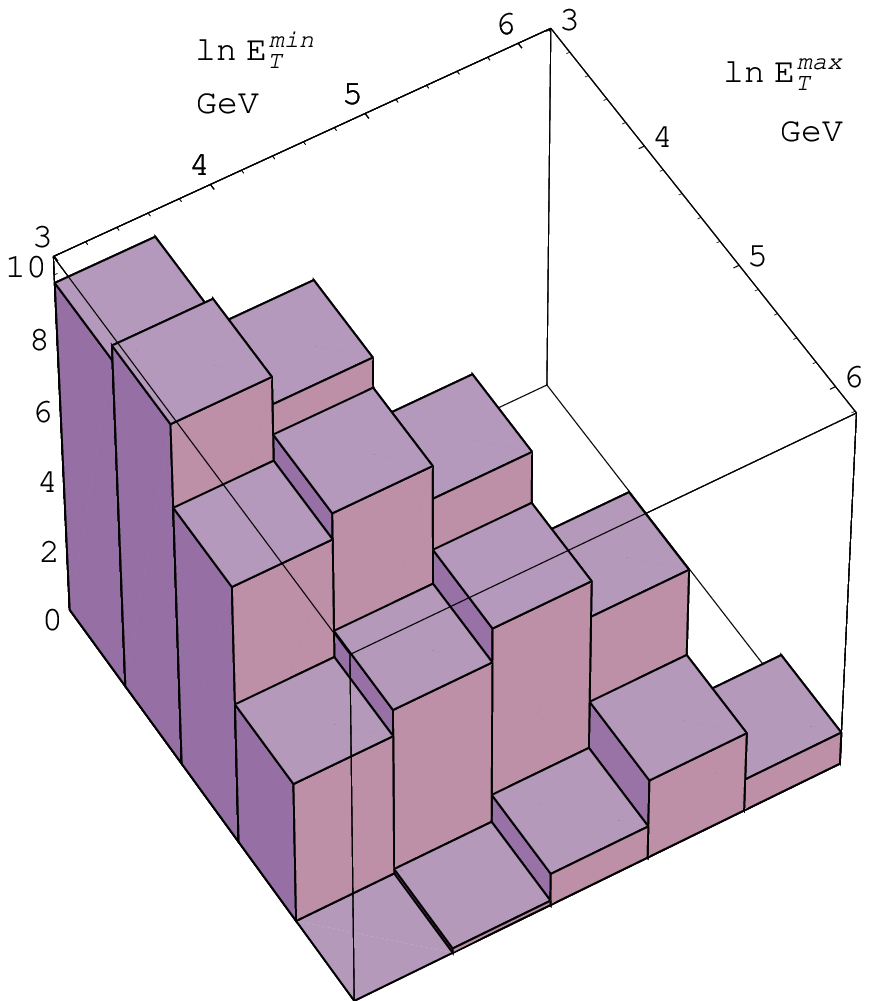,width=0.44\textwidth,clip=}
   \hfill
   \epsfig{figure=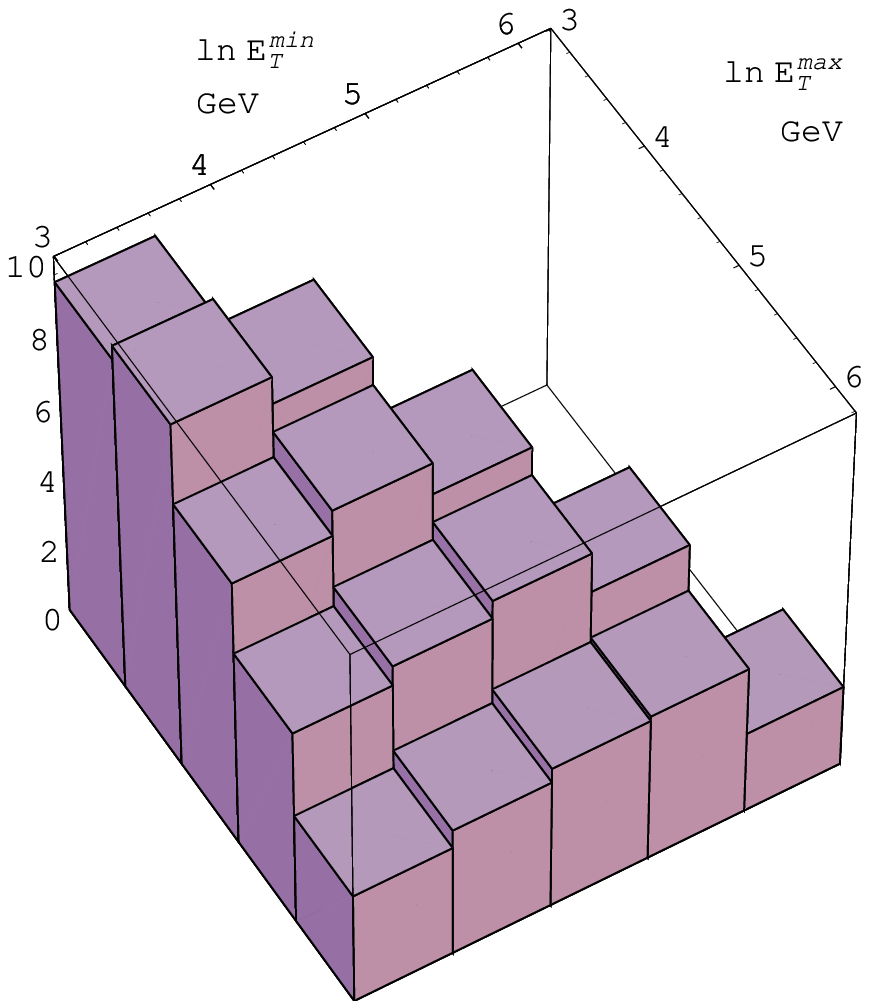,width=0.44\textwidth,clip=} }
\ccaption{}{ \label{fig:AnomC}
   Cross section $\sigma$~[fb/10] for $W$-pair production at the
   Tevatron with no anomalous couplings (left) and for the HISZ
   scenario with $\alpha_W = \alpha_{W\phi} = \alpha_{B\phi} = 0.1$
   (right). Standard cuts have been applied and the scale has been set
   to $\mu = \mu_{\rm st}$ as defined in \eqn{must}. 
}
\end{figure}

The same results are also shown in Fig.~\ref{fig:AnomC}, where we plot
the natural logarithm of the total cross section (in units of $10^{-1}$~fb)
for each bin for the Standard Model with $\mu=\mu_{\rm st}$ (left) and
the HISZ scenario with $\mu=\mu_{\rm st}$ (right). Again, the
significant differences in the high $E_T$ bins become apparent.

\section{Conclusions}
\label{ConclusionsSection}

We have presented a general purpose Monte Carlo program that is able
to compute any infrared-safe quantity in vector boson pair production at
hadron colliders at next-to-leading order in the strong coupling
constant $\as$. This program generalizes previous calculations
(with the exception of the recent ref.~\cite{CE}) in that the spin 
correlations are fully taken into account. The decay of the vector bosons 
into leptons was included in the narrow-width approximation, whereas
{\tt MCFM}~\cite{CE} also includes the singly-resonant diagrams needed to 
go beyond this approximation.  For the total cross-section, computed in
the narrow-width approximation, our results agree perfectly with those of 
{\tt MCFM} for the same choice of input parameters.

As an illustration of the usefulness of the program we presented several
distributions for the Run~II at the Tevatron and the LHC. However, we
refrained from performing a detailed phenomenological analysis;
this is probably best done once the data are available. 

In addition to Standard Model processes, we considered also the
inclusion of anomalous couplings between the vector bosons. We
presented the one-loop amplitudes for a generalized trilinear
$W^+W^-Z$ and $W^+W^-\gamma$ vertex. The inclusion of these amplitudes
into our program allows the calculation of anomalous effects at
next-to-leading order, with full spin correlations.  These
effects are shown to be very prominent for large transverse momentum
of the gauge bosons, confirming results of ref.~\cite{BHOWW}.  
Such an analysis at Run II of the Tevatron should yield improved 
bounds on anomalous couplings, although for large improvements one 
probably has to wait for the LHC~\cite{EllisonWudka}.

\bigskip

{\large \bf Acknowledgments}
\bigskip

L.D. and A.S. would like to thank the Theory Group of ETH Z\"urich for
its hospitality while part of this work was carried out.  We are
grateful to John Campbell and Keith Ellis for assistance in the
comparison of our results with those of ref.~\cite{CE}.


\def\np#1#2#3  {{\it Nucl. Phys. }{\bf #1} (19#3) #2}
\def\nc#1#2#3  {{\it Nuovo. Cim. }{\bf #1} (19#3) #2}
\def\pl#1#2#3  {{\it Phys. Lett. }{\bf #1} (19#3) #2}
\def\pr#1#2#3  {{\it Phys. Rev. }{\bf #1} (19#3) #2}
\def\prl#1#2#3  {{\it Phys. Rev. Lett. }{\bf #1} (19#3) #2}
\def\prep#1#2#3 {{\it Phys. Rep. }{\bf #1} (19#3) #2}
\def\zp#1#2#3  {{\it Z. Phys. }{\bf #1} (19#3) #2}
\def\rmp#1#2#3  {{\it Rev. Mod. Phys. }{\bf #1} (19#3) #2}
\def\mpl#1#2#3 {{\it Mod. Phys. Lett. }{\bf #1} (19#3) #2}

\end{document}